\documentclass[twoside,reqno]{bjp}

\usepackage{graphicx}
\usepackage{amssymb,amsmath}
\usepackage{cite}
\usepackage{times}
\usepackage{bm}

\newcommand{\tc}{\ensuremath{T_\mathrm{c}}}
\newcommand{\Fref}[1]{Fig.~\ref{#1}}
\newcommand{\Eqref}[1]{Eq.~(\ref{#1})}
\newcommand{\be}{\begin{equation}}
\newcommand{\ee}{\end{equation}}

\begin{document}

\setcounter{page}{106} 

\sloppy 
\raggedbottom

\title{Correlation between {\slshape{T}}$_{\bm{\textsf{c}}}$  
and the Cu~4s level reveals the 
mechanism of high-temperature superconductivity%
\thanks{This work is dedicated, 
in Memoriam Academician Professor Matey Mateev.}
}

\runningheads{Correlation between $\mathrm{T}_c$
and the Cu~4s level reveals  the 
mechanism\dots}{Z.~D.~Dimitrov~\textit{et al.}}

\begin{start}
\author{Z.~D.~Dimitrov}{}, 
\coauthor{S.~K.~Varbev}{},
\coauthor{K.~J.~Omar}{},
\coauthor{A.~A.~Stefanov}{},\\
\coauthor{E.~S.~Penev%
\thanks{Present address: Department of Mechanical Engineering and 
Materials Science, Rice University, Houston, Texas, USA}}{},
\coauthor{T.~M.~Mishonov%
\thanks{Corresponding author: \texttt{tmishonov@phys.uni-sofia.bg}}}{}

\address{Department of Theoretical Physics, Faculty of Physics,
St.~Clement of Ohrid University at Sofia, 5 James Bourchier Blvd.,
BG-1164 Sofia, Bulgaria}{}

\received{16 March 2011}

\begin{Abstract}
Band structure trends in hole-doped cuprates and correlations with
\tc\  are interpreted within the s--d exchange mechanism
of high-\tc\ superconductivity. The dependence of \tc\ on the
position of the copper 4s level finds a natural explanation in the
generic Cu~3d, Cu~4s, O~2p$_\textit{x}$ and O~2p$_\textit{y}$ four-band model. The
Cu~3d--Cu~4s intra-atomic exchange interaction is incorporated in
the standard BCS scheme. This dependence of \tc\ in the whole
interval of 25--125~K has no alternative explanation at present, and
possibly this quarter of a century standing puzzle is already solved.
\end{Abstract}

\end{start}

\section{Introduction}

Quarter of a century after the dramatical discovery of high
temperature superconductivity (HTS) the problem of its mechanism
remains one of the longest-standing puzzles in the history of science.
The cornucopia of ideas is immense -- almost every quantum process was
tested whether it is the long sought mechanism of HTS. In parallel, the
intensive experimental research made cuprates into the most
investigated materials. First-principles electronic structure
calculations play an important role in understanding the physics of
these materials. The Fermi surface of these highly anisotropic
crystals is almost cylindrical with rounded-square cross-section.
Fifteen years after the beginning of the cuprates era the eye of the
professionalist has uncovered a subtle correlation~\cite{Pavarini:01}
between the shape of the Fermi contour and the critical temperature
\tc\ for optimally hole doped cuprates. Pavarini \textit{et al.}~\cite{Pavarini:01}
considered many hole doped cuprates: 
Ca$_2$CuO$_2$Cl$_2$,
La$_2$CuO$_4$,
Bi$_2$Sr$_2$Cu$_2$O$_6$,
Tl$_2$Ba$_2$CuO$_6$,
Pb$_2$Sr$_2$Cu$_2$O$_6$,
TlBaLaCuO$_5$,
HgBa$_2$CuO$_4$,
LaBa$_2$Cu$_3$O$_7$,
La$_2$CaCu$_2$O$_6$,
Pb$_2$Sr$_2$YCu$_3$O$_8$,
YBa$_2$Cu$_3$O$_7$,
Tl$_2$Ba$_2$CaCu$_2$O$_8$,
Tl$_2$Ba$_2$Ca$_2$Cu$_3$O$_{10}$,
HgBa$_2$Ca$_2$Cu$_3$O$_8$,
HgBa$_2$CaCu$_2$O$_6$;
the list can be extended but we believe that a universal correlation was discovered.

In an acceptable approximation the electronic band structure can be
described by the four-band Linear Combination of Atomic Orbitals (LCAO) model with on-site energies
$\epsilon_\mathrm{s}, \epsilon_\mathrm{p}, \epsilon_\mathrm{d}$, and hopping parameters $t_\mathrm{sp},
t_\mathrm{pd}, t_\mathrm{pp}.$ In this approximation the Hilbert space spans over
the Cu~3d$_{x^2-y^2}$, Cu~4s, O~2$p_x$ and O~2$p_y$ valence
orbitals. The range parameter $r\equiv1/2(1+s)$ which determines the shape
of the Fermi contour is determined by the LCAO parameters and the
Fermi energy $E_\mathrm{F}$
\be
s(E_\mathrm{F})=(\epsilon_\mathrm{s}-E_\mathrm{F})(E_\mathrm{F}-\epsilon_\mathrm{p})/(2t_\mathrm{sp})^2.
\ee

\begin{figure}[t]
\centering
\includegraphics[height=6cm]{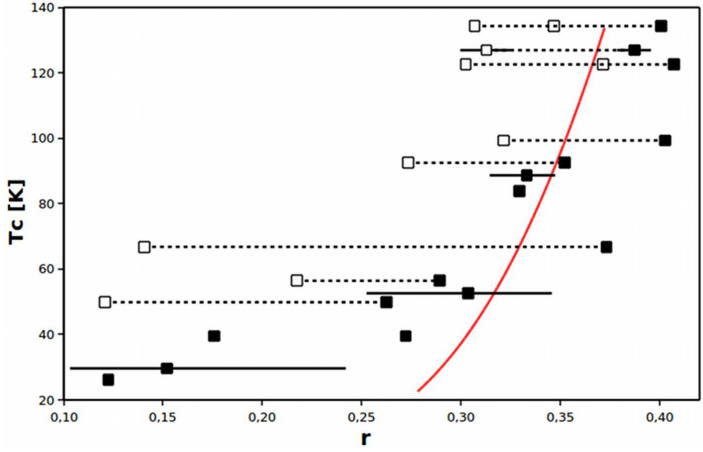}
\caption{Critical temperature of hole doped cuprates versus band
  structure parameter $r$ by Pavarini~\textit{et al.}~\cite{Pavarini:01}.
The theoretical curve is calculated according to \Eqref{BCS} (see text for details).}
\label{fig_crucial}
\end{figure}

The correlation between \tc\ and $r$ from the work by Pavarini
\textit{et al.}~\cite{Pavarini:01} is reproduced in
\Fref{fig_crucial}. How such a correlation can possibly exist?
Superconductivity is, certainly, created by some interaction, whilst
the electronic band structure describes the properties of independent
electrons. Is it then possible to uncover an unknown mechanism of
interaction investigating only the properties of noninteracting
particles? The \tc--$r$ correlation covers the whole temperature range
of HTS, but now ten years it remains unexplained. We suppose that this
correlation is as important for HTS, as was the isotope effect for the
conventional phonon superconductors half a century ago. The aim of the
present work is to interpret the correlation reported by
Pavarini~\textit{et al.}~\cite{Pavarini:01} in the framework of some of
the models of HTS.
\section{{\slshape{T}}$_{\bm{\textsf{c}}}$--{\slshape{r}} correlation within the s--d theory}

Let us introduce a small modification of the $s$ parameter,
$\tilde{s}=\displaystyle \frac{t_\mathrm{sp}}{t_\mathrm{pd}}s.$ Its reciprocal value 
\be
\frac{1}{\tilde{s}(\epsilon)} =
\frac{4t_\mathrm{pd}t_\mathrm{sp}}{(\epsilon_\mathrm{s}-\epsilon)(\epsilon-\epsilon_\mathrm{p})},
\label{eq:stilde}
\ee
has an energy denominator typical for the perturbation theory 
as applied to the secular equation of the generic four-band model
\begin{equation}
\begin{pmatrix}
\epsilon_\mathrm{d} & 0 & t_\mathrm{pd}s_x &  -t_\mathrm{pd}s_y \\
0 & \epsilon_\mathrm{s}  & t_\mathrm{sp}s_x & t_\mathrm{sp}s_y \\
 t_\mathrm{pd}s_x  &  t_\mathrm{sp}s_x  & \epsilon_\mathrm{p}  &  t_\mathrm{pp}s_x s_y  \\
-t_\mathrm{pd}s_y & t_\mathrm{sp}s_y & -t_\mathrm{pp}s_x s_y & \epsilon_\mathrm{p}
\end{pmatrix}
\begin{pmatrix}
 D_\mathbf{p} \\ S_\mathbf{p} \\ X_\mathbf{p} \\ Y_\mathbf{p}
\end{pmatrix}
=\epsilon_\mathbf{p} 
\begin{pmatrix}
 D_\mathbf{p} \\ S_\mathbf{p} \\ X_\mathbf{p} \\ Y_\mathbf{p}
\end{pmatrix},
\end{equation}
where $\epsilon_\mathbf{p}$ is the electron energy for the conducting
d-band, $s_x=2\sin(p_x/2),$  $s_y=2\sin(p_y/2),$ and the wave function is normalized
$D^2_\mathbf{p}+S^2_\mathbf{p}+X^2_\mathbf{p}+Y^2_\mathbf{p}=1.$
The approximate solution for small hopping amplitudes 
\begin{equation}
\begin{pmatrix}
 D_\mathbf{p} \\ S_\mathbf{p} \\ X_\mathbf{p} \\ Y_\mathbf{p}
\end{pmatrix}
\approx
\left(
\begin{array}{c}
  1 \\ 
\left(s_x^2-s_y^2\right) / 4\tilde{s}(\epsilon)\\
  \displaystyle\frac{t_\mathrm{pd}}{\epsilon-\epsilon_\mathrm{p}}s_x\\ 
  -\displaystyle
\frac{t_\mathrm{pd}}{\epsilon-\epsilon_\mathrm{p}}s_y
\end{array}\right),
\end{equation}

has a simple interpretation: in the initial approximation we have a
pure Cu~3d state with $D_\mathbf{p}\approx 1,$ in first
approximation we have a linear dependence from the $t_\mathrm{pd}$ amplitude
on O~2p levels $X_\mathbf{p},\; Y_\mathbf{p} \propto t_\mathrm{pd}$, and
finally in second approximation, the amplitude on the Cu~4s orbital
$S_\mathbf{p}\propto t_\mathrm{pd}t_\mathrm{sp}$ is included by the second virtual
transition proportional to $t_\mathrm{sp}$. In short, the Cu~4s amplitude of
the conduction band can be phrased as 3d--to--4s--by--2p.

As it was concluded by Pavarini~\textit{et al.}~\cite{Pavarini:01}
that materials with lower $\epsilon_\mathrm{s}$ tend to be those with higher
observed values of $T_{\mathrm{c}\,\,max}$. In the materials with
higher $T_{\mathrm{c}\,\,max}$ the axial orbital is almost Cu~4s. What is the
simplest interpretation of these observations? There is an emerging
consensus that superconductivity in cuprates is created by some
exchange interaction. The Pavarini \textit{et al}~\cite{Pavarini:01}
correlation gives that higher $T_{\mathrm{c}\,\,\mathrm{max}}$ is determined by
the highest Cu~4s amplitude $S_\mathbf{p}$.  Now we have to recall that
the most usual 3d--4s intra-atomic exchange has one of the
largest amplitudes in condensed matter physics. 
We also have to point out that transition metal ion-ion exchange integrals $J_{ii},$ 
which create antiferromagnetism in insulator phase 
are smaller than intra atomic exchange integral $J_{sd}.$

\begin{figure}[t]
\centering
\includegraphics[height=7cm]{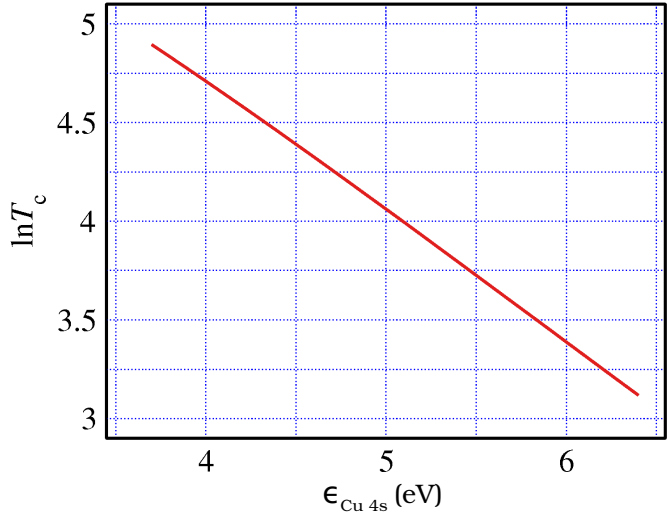}
\caption{Logarithm of \tc\ (K) versus $\epsilon_{\text{Cu}\,4\mathrm{s}}$. Science
  starts with simplicity -- the almost linear behavior corresponds to
  the well-known BCS formulae given by Eqs.~(\ref{BCSformula}) and (\ref{lnTc}).}
\label{fig_logTc}
\end{figure}

The intensive investigations of the Kondo effect have demonstrated
that the anti-ferromagnetic sign of the two electrons s--d
exchange is the rule and the ferromagnetic sign is an exception.
Incorporated in the BCS scheme 
for the calculation of the order parameter $\Xi,$ 
the anti-ferromagnetic sign gives
pairing in singlet channel with momentum dependent gap 
\be
\Delta_\mathbf{p}(T)=\Xi(T)\chi_\mathbf{p}, 
\qquad \chi_\mathbf{p} \equiv S_\mathbf{p}D_\mathbf{p}.  
\ee 
This gap is included in the fermion excitation energy
\be E_{\mathbf{p}} \equiv (\eta_{\mathbf{p}}^2 +
\Delta^2_{\mathbf{p}})^{1/2},\qquad \eta_{\mathbf{p}} \equiv
\epsilon_\mathbf{p} -E_\mathrm{F}, 
\ee
and for the temperature dependent order parameter $\Xi(T)$ we have the
standard BCS equation
\be
\label{BCS}
2J_\mathrm{sd} \left<\frac{\chi_{\mathbf{p}}^2}{2E_{\mathbf{p}}} 
\tanh\left(\frac{E_{\mathbf{p}}}{2k_\mathrm{B} T}\right) \right> = 1, 
\quad 
\left<f_\mathbf{p}\right> \equiv \int_{0}^{2\pi} 
\int_{0}^{2\pi} \frac{dp_x \, dp_y}{(2\pi)^2}f(\mathbf{p})
\ee
where $\langle\dots\rangle$ denotes momentum--space averaging over the
Brillouin zone. For a pedagogical derivation of the BSC gap equation
in the present notations see the textbook~\cite{Mishonov:11} and references therein.
Slightly below the critical temperature where the order parameter
disappears $\Xi(\tc -0)=0$ the gap equation reads
\be
\label{Tc}
\quad \left<\frac{\chi_{\mathbf{p}}^2}{\eta_{\mathbf{p}}} 
\tanh\left(\frac{\eta_{\mathbf{p}}}{2k_\mathrm{B} \tc}\right) \right> 
= \frac{1}{J_\mathrm{sd}}.
\ee
Supposing that $J_\mathrm{sd}$, being an intra-atomic process, is weakly
temperature dependent we can determent its value using band parameters
$\epsilon_\mathrm{s},$ $\epsilon_\mathrm{d},$ $\epsilon_\mathrm{p},$ $t_\mathrm{sp},$ $t_\mathrm{pd},$
$t_\mathrm{pp},$ $E_\mathrm{F}$ and known \tc.  Then we can calculate, for
example, the dependence of $T_{\mathrm{c}\,\,\mathrm{max}}$ by position of the Cu~4s
levels $\epsilon_\mathrm{s}$.  One illustrative example is shown in
\Fref{fig_logTc}. The almost linear dependence has to have a simple
qualitative interpretation. Let us try to reveal this simplicity using
the BCS interpolation formula 
\be k_\mathrm{B} T_\mathrm{c} =
1.14\,\hbar\omega_\mathrm{D}\mathrm{e}^{-1/N(0)V}.
\label{BCSformula}
\ee
In the present case $\omega_\mathrm{D}$ is an energy parameter of
order of the bandwidth, $N(E_\mathrm{F})$ is electronic density of states per
spin at the Fermi level and
\be 
V = J_\mathrm{sd}
\frac{4t_\mathrm{pd}t_\mathrm{sp}}{(\epsilon_\mathrm{s}-E_\mathrm{F})(E_\mathrm{F}-\epsilon_\mathrm{p})}
= J_\mathrm{sd}/\tilde{s}(E_\mathrm{F}).
\ee
The interpolation BCS formula gives 
\be 
\label{lnTc}
\ln(\tc) \approx \ln(1.14\omega_\mathrm{D}) -
\tilde{s}(E_\mathrm{F})/[N(E_\mathrm{F})J_\mathrm{sd}], 
\ee 
which according to \Eqref{eq:stilde}
correlations reported by Pavarini~\textit{et al.}~\cite{Pavarini:01}
are actually correlations between the critical temperature
$T_{\mathrm{c}\,\,\mathrm{max}}$ and the BCS coupling constant
$J_\mathrm{sd}/\tilde{s}(E_\mathrm{F})$. This general correlation is
typically perturbed by stripes, inhomogeneities, magnetic phenomena,
structural phase transition, apex oxygen, chemical substitution, and
many other accessories of HTS cuprates. Nevertheless, they can be
clearly seen for all hole doped cuprates. This qualitative agreement
gives hope that four-band model with incorporated s--d exchange can
become a standard model for superconductivity of cuprates. The key
role of the $\tilde{s}(E_\mathrm{F})$ parameter reveals why cuprates
are unique for reaching high-\tc. Imagine that we can easily tune the
position of the Cu~4s level $\epsilon_\mathrm{s}$.  The pre-exponential
factor $\omega_\mathrm{D}$ is so big that we can easily reach a
\emph{sauna-temperature} superconductivity if $\epsilon_\mathrm{s}$ is small
enough. Maximal \tc\ is reached upon 3d--2p--4s hybridization.
For the Cu--O duet we have maximal triple coincidence of levels which
ensures the success of the CuO$_2$ plane.  For all other combinations
of transition metal with a chalcogenide $\tilde{s}$ is much bigger.
For 25 years many ways to decrease $\epsilon_\mathrm{s}$ were
empirically found: apex oxygen, bilayer hopping, pressure, etc.
Perhaps only metastable artificial layers are not completely
investigated.

\section{Computational method}

The parent CuO$_2$ layer is an insulator with a half-filled conduction
band. Doping with $\tilde{p}$ holes per Cu atom results in metalization
with hole filling $f=\frac{1}{2}+\tilde{p}.$ The optimal doping
corresponds to $\tilde{p}_\mathrm{max}=0.16$ and
$f_\mathrm{max}=0.66$.  We will use 66\% hole filling for all examples
in the present work. The Fermi level is determined by the condition
\be
f=\left<\theta(\epsilon_\mathbf{p}>E_\mathrm{F})\right>,\quad \mbox{where}\quad
\theta(\epsilon_\mathbf{p}>E_\mathrm{F}) = 
\begin{cases}
  1 & \text{if } E_\mathbf{p} > E_\mathrm{F},\\
  0 & \text{if } E_\mathbf{p} < E_\mathrm{F}.
\end{cases}
\ee

The parameters of the four-band model can be determined by comparison
with first-principles electronic structure calculations.  For example,
using the $\Gamma$ point one can determine the on-site energies
$\epsilon_\mathrm{d}=\epsilon(p_x=0,p_y=0)$ for the conduction band.
In the present paper we use as an illustration a set of parameters (in eV)
similar to Ref.~\cite{Pavarini:01}
\begin{gather} 
  \epsilon_\mathrm{s}=5.4,\quad \epsilon_\mathrm{p}=-1, \quad \epsilon_\mathrm{d}=0,\\ 
  t_\mathrm{sp}=2,\quad t_\mathrm{pd}=1.5, \quad t_\mathrm{pp}=0.2 \nonumber.
\end{gather} 
eV Assuming $T_\mathrm{c,\,\,max}=90$~K for $\epsilon_\mathrm{s}=5.4$
we obtain, according to \Eqref{Tc}, $J_\mathrm{sd}=2.44\;\mathrm{eV}.$
Then for so fixed $J_\mathrm{sd}$ we can calculate the correlations of
$T_\mathrm{c,\,\,max}$ with the range parameter
\be 
r\equiv \frac{1}{2(1+s)}. 
\ee 

Let us discuss briefly our finding.
Within the four-band model one can derive an exact equation for
constant energy contours
\be
-2t(\epsilon)[\cos(p_x)+\cos(p_y)] +4t'(\epsilon)\cos(p_x)\cos(p_y)
=q(\epsilon),
\ee
where $t(\epsilon),$ $t'(\epsilon)$ and $q(\epsilon)$ are polynomial
functions of energy. The ratio of the effective intra-layer hopping
parameters
\be \frac{t'(E_\mathrm{F})}{t(E_\mathrm{F})}\propto
r(E_\mathrm{F}), 
\ee 
describes small variations of the shape of the Fermi contour
influenced by the position of the Cu~4s level $\epsilon_\mathrm{s}$.
\section{Conclusions}
Electronic structure experts should be proud that after many years of
systematic research band calculations have revealed that \tc\ depends
on the Cu~4s level.  This numerical experiment is actually the
crucial one for understanding the mechanism of HTS. The $r$ parameter
is introduced in electronic structure calculations in such a way, that
for the first time we have a mechanism of a physical phenomenon
possibly revealed by computer.

We advocate a conventional theoretical explanation of the \tc-$r$
correlations~\cite{Pavarini:01} which has no alternative among other
theories of HTS.  We have to wait for the appearance of some other
descriptions, because this important correlation covers a hole range
of HTS. The LCAO model was only a tool for the theoretical analysis
of the mechanism of HTS. It should be interesting to determine
$S_\mathbf{p}$ and $D_\mathbf{p}$ directly from partial wave analysis
at the muffin spheres directly by the electronic band calculations.
The interpolating $\chi_\mathbf{p}=S_\mathbf{p} D_\mathbf{p}$ function
can be directly substituted in the BCS gap equation (\ref{BCS}) and
the equation for \tc, \Eqref{Tc}.  For the cuprates $\chi_\mathbf{p}$
well describes the experimental data for the gap anisotropy~$\Delta_\mathbf{p}$ 
on the Fermi surface. It remains to be seen if
this mechanism is applicable only to cuprates.  

The iron-based pnictides are suspected to pursue an Oscar for
supporting role.  It will be extremely interesting to probe if
\Eqref{Tc} can describe the common trends for the \tc\ and gap
anisotropy of ferro-pnictides. In this regards, finally we would like
to re-analyze the ``boost'' that the physics of superconductivity got
from the iron pnictides. Very recently, Mazin~\cite{mazin10} recalled the
Matthias' rules, well known in the physicists’ folklore: 1) A high
symmetry is good; cubic symmetry is the best.  2) A high density of
electronic states is good.  3) Stay away from oxygen.  4) Stay away
from magnetism.  5) Stay away from insulators.  6) Stay away from
theorists.

In order to emphasize the common properties of cuprates and iron
pnictides we feel it compelling to rephrase these rules: (1) A high
symmetry is good, square symmetry is the best, layered structure
ensures empty s-band.  (2) Having empty s-band, the Fermi level falls
in the narrow d-band which ensures high density of states. (3) The
oxygen and pnictide p-orbitals are perfect ``go-between'' for the
transition metal 3d and 4s orbitals. These oxidants create significant
4s polarization of the conduction 3d band. (4) We are just in the
epicenter of magnetism; the s-d exchange amplitude which creates the
ferromagnetism of iron, possibly creates the superconductivity of iron
pnictides and cuprates.  (5) Having narrow 3d band in the case of
half-filling we observe metal-insulator transition. Transition to
insulator phase is a hint for high density of states at the Fermi
level for a doped compound.  (6) Favorites of great socialists like
Lenard, Kapica and Stark, have given similar advice. In order to
surmount this weakness it is necessary to remedy the inferiority
complex of the band calculators. One Nobel prize is a good initial
dose of this treatment. Up to now there is no Nobel prize awarded to
computational physics, but the considered in this paper $T_c$-$r$
correlations are a typical physical phenomenon whose nature is
revealed by numerical calculations. Prediction of an artificial
structure with high $\tilde{s}$ parameter, further synthesis by
solid-state chemists and measurement of $T_c$ may successfully
initiate a new direction of intensive research.

In conclusion, we briefly re-state the main property of high-$T_c$
superconductivity---the triple coincidence of the d-, s-, and p-levels
which ensures big enough BCS coupling constant. This triple
coincidence is analogous to the sequence of vernal equinox, full moon,
and Sunday (the end of the winter season, the end of the moon month
and the end of the week): a triple holiday creating the Great-day. In
this sense the high-$T_c$ superconductivity is the Easter of condensed
matter physics.

The authors are thankful to Prof.~Ivan Zhelyazkov for a critical
reading of the manuscript and to Alvaro de R\'ujula for his interest and
comments during the scientific conference in memory of Matey Mateev,
held in Sofia, 11--12 April 2011.


\begin{thebibliography}{2}
\bibitem{Pavarini:01} 
  E.~Pavarini, I.~Dasgupta, T.~Saha-Dasgupta,
  O.~Jepsen, and O.~K.~Andersen, Band-Structure Trend in Hole-Doped
  Cuprates and Correlation with $T_{c\,\mathrm{max} }$, (2001)
  \textit{Phys. Rev. Lett.} \textbf{87} 047003.

\bibitem{Mishonov:11} T.~M.~Mishonov and E.~S.~Penev (2011)
  \textit{Theory of High Temperature Superconductivity: A Conventional
    Approach}, World Scientific, Singapore, Chap. 2.

\bibitem{mazin10} I.~Mazin, Superconductivity gets
  an iron boost, (2010) \textit{Nature} \textbf{464}, 183--186.

\end{thebibliography}
\end{document}